\documentclass{aastex}
\usepackage{emulateapj5}
\usepackage{onecolfloat5}
\usepackage{apjfonts}

\def\plotone#1{\centering \leavevmode
\epsfxsize= 1.0\columnwidth \epsfbox{#1}}

\def\gsim{\;\rlap{\lower 2.5pt
 \hbox{$\sim$}}\raise 1.5pt\hbox{$>$}\;}
\def\lsim{\;\rlap{\lower 2.5pt
   \hbox{$\sim$}}\raise 1.5pt\hbox{$<$}\;}
\newcommand{\be}{\begin{equation}}
\newcommand{\beq}{\begin{equation}}
\newcommand{\ba}{\begin{eqnarray}}
\newcommand{\ee}{\end{equation}}
\newcommand{\eeq}{\end{equation}}
\newcommand{\ea}{\end{eqnarray}}

\usepackage{natbib}

\begin{document}
\twocolumn[
\submitted{Submitted to ApJ Letters}
\title{Was Star--Formation Suppressed in High-Redshift Minihalos?}

\author{Zolt\'an Haiman and Greg L. Bryan}

\affil{Department of Astronomy, Columbia University, 550 West 120th Street, New York, NY 10027}

\begin{abstract}
The primordial gas in the earliest dark matter halos, collapsing at
redshifts $z\sim 20$, with masses $M_{\rm halo}\sim 10^6~{\rm
M_\odot}$, and virial temperatures $T_{\rm vir}< 10^4$K, relied on the
presence of molecules for cooling. Several theoretical studies have
suggested that gas contraction and star--formation in these minihalos
was suppressed by radiative, chemical, thermal, and dynamical feedback
processes.  The recent measurement by the {\it Wilkinson Microwave
Anisotropy Probe} ({\it WMAP}) of the optical depth to electron
scattering, $\tau\approx 0.09\pm 0.03$, provides the first empirical
evidence for this suppression.  The new {\it WMAP} result is
consistent with vanilla models of reionization, in which ionizing
sources populate cold dark matter (CDM) halos down to a virial
temperature of $T_{\rm vir}=10^4$K.  On the other hand, we show that
in order to avoid overproducing the optical depth, the efficiency for
the production of ionizing photons in minihalos must have been about
an order of magnitude lower than expected {\it and} lower than the
efficiency in large halos that can cool via atomic hydrogen ($T_{\rm
vir} > 10^4$K).  This conclusion is insensitive to assumptions about
the efficiency of ionizing photon production in the large halos, as
long as reionization ends by $z=6$, as required by the spectra of
bright quasars at $z\lsim 6$. Our conclusion is strengthened if the
clumping of the ionized gas evolves with redshift, as suggested by
semi--analytical predictions and three-dimensional numerical
simulations.\\
\end{abstract}]


\section{Introduction}
\label{sec:introduction}

How and when the intergalactic medium (IGM) was reionized is one of
the long outstanding questions in astrophysical cosmology, holding
clues about the onset of structure formation in cold dark matter (CDM)
cosmologies, and the nature of the first generation of light sources
(see Barkana \& Loeb 2001 for a review).  The reionization history can
be probed via the scattering of the cosmic microwave background (CMB)
photons by the free electrons liberated during reionization (see,
e.g., Haiman \& Knox 1999 for a review).  The total electron
scattering optical depth $\tau=0.088^{+0.028}_{-0.034}$ (Spergel et
al. 2005, Table 5) that has been inferred from the {\it WMAP}
three--year polarization data (Page et al. 2005) implies that
reionization occurred relatively recently, at $z_r\lsim
11$.\footnote{Assuming an abrupt transition from a fully neutral to a
fully ionized IGM, and assuming helium is once ionized at the same
redshift as hydrogen, we find $z_r < 10.9^{+2.2}_{-3.1}$ for the
cosmological parameters listed in Table 5 in Spergel et al. (2005).}

An outstanding question, and the focus of this {\it Letter}, is
whether the gas in the majority of the first--generation low--mass
halos was able to cool and form ionizing sources (stars and/or black
holes [BHs]), and contribute to reionizing the IGM.  The neutral gas
in halos with virial temperatures $T_{\rm vir}\lsim 10^4$K (hereafter
``minihalos'') cannot cool via atomic hydrogen alone, and the presence
of ${\rm H_2}$ (or other) molecules is a necessary condition for the
gas to reach high densities and to ultimately form stars or BHs in
these halos (Saslaw \& Zipoy 1967).  As a result, these halos are
distinct from larger halos with $T_{\rm vir}\gsim 10^4$K, which are
expected to have different star--formation properties (Oh \& Haiman
2002; Johnson \& Bromm 2006). Since the smallest halos are the first
to collapse, whether or not typical minihalos are able to host
ionizing sources is the most significant factor in determining when
reionization began, and driving the value of $\tau$ (Haiman \& Holder
2003 [HH03]; see Abel \& Haiman 2000 for a review focusing on the role
of ${\rm H_2}$ molecules for reionization).

In isolation, minihalos with virial temperatures as low as a few $100$
K could form enough ${\rm H_2}$, via gas--phase chemistry, for
efficient cooling and gas contraction (Haiman, Thoul \& Loeb 1996;
Tegmark et al. 1997).  However, ${\rm H_2}$ molecules are fragile, and
can be dissociated by soft UV radiation absorbed in their Lyman-Werner
(LW) bands (e.g. Haiman, Rees \& Loeb 1997; Ciardi et al 2000;
Ricotti, Gnedin \& Shull 2001).  In patches of the IGM corresponding
to fossil HII regions that have recombined, the gas retains excess
entropy, which can reduce gas densities in the cores of collapsing
halos, and lower the LW radiation background that photodissociates
${\rm H_2}$ (Oh \& Haiman 2002; Gnedin 2000).  On the other hand,
positive feedback effects, such as the presence of extra free
electrons (beyond the residual electrons from the recombination epoch)
from protogalactic shocks (Shapiro \& Kang 1987; Ferrara 1998), from a
previous ionization epoch (Oh \& Haiman 2003; Susa et al. 1998), or
from X--rays (Haiman, Rees \& Loeb 1996, Oh 2001; Ricotti et
al. 2002a,b), can enhance the ${\rm H_2}$ abundance.  Whether or not
gas cooling in minihalos was efficiently quenched globally has
remained unclear, with numerical simulations generally favoring less
quenching (Machacek et al. 2001, 2003; Ricotti et al. 2002a,b; Kuhlen
\& Madau 2005) than predicted in semi--analytical models.

The purpose of this {\it Letter} is to place constraints on the
efficiency of minihalos to produce and inject ionizing photons into
the IGM, using the value of $\tau$ inferred from the {\it WMAP}
polarization data.  Although the absolute values of the relevant
efficiency parameters are uncertain, there is a robust lower limit on
their combination from the requirement that reionization is completed
prior to $z=6$. As a result, a measurement of $\tau$ permits robust
conclusions about the evolution of the efficiency with redshift or
halo mass--scale (HH03; Cen 2003; Onken \& Miralda-Escud\'e 2004).
The main conclusion of this {\it Letter} is that the efficiency with
which minihalos injected ionizing radiation into the IGM was about an
order of magnitude lower than expected {\it and} lower than in larger
halos, requiring negative feedback at high redshift.

Throughout this letter, we adopt the background cosmological
parameters as measured recently by the {\it WMAP} experiment (Spergel
et al. 2005, Table 5), $\Omega_m=0.24$, $\Omega_{\Lambda}=0.76$,
$\Omega_b=0.0407$, $h=0.72$ and an initial matter power spectrum $P(k)
\propto k^n$ with $n=0.96$ and normalization $\sigma_8=0.76$. The
values of these parameters determine the required overall
efficiencies, but do not otherwise have a significant effect on our
conclusions.

\section{Models of Reionization}
\label{sec:models}

In this section, we briefly describe our semi--analytical model of the
reionization process.  The treatment follows the prescriptions in
HH03, and the reader is referred to this paper for a detailed
description; we only recapitulate the main features and a few
important modifications here.  The models track the total
volume--filling factor $F_{\rm HII}(z)$ of ionized regions, assuming
that ionizing sources are located inside virialized dark matter
halos. The sources create ionized HII regions, which expand into the
IGM at a rate dictated by the source luminosity, and by the background
gas density and clumping factor.\footnote{We do not explicitly treat
the reionization of helium, but assume the HeII/HeI fraction follows
the HII/HI fraction when we compute $\tau$.  For neutral/doubly
ionized helium, $\tau$ would be smaller/larger by $\sim 8\%$.}

The efficiency with which the ionizing source(s) associated with each
halo inject ionizing photons into the IGM is parameterized by the
product $\epsilon \equiv N_\gamma f_* f_{\rm esc}$, where $f_* \equiv
M_*/(\Omega_{\rm b}M_{\rm halo}/\Omega_m)$ is the fraction of baryons
in the halo that turns into stars; $N_\gamma$ is the mean number of
ionizing photons produced by an atom cycled through stars, averaged
over the initial mass function (IMF) of the stars; and $f_{\rm esc}$
is the fraction of these ionizing photons that escapes into the IGM.
We define separate efficiencies, $\epsilon_{\rm mini}$, $\epsilon_{\rm
large}$ for the minihalos and large halos, respectively.  The
distinction between the halos is made at $T_{\rm vir}=10^4$K, with the
relation $T_{\rm vir}\approx 1800 (M/10^6{\rm M_\odot})^{2/3}
(1+z)/21$ K.  We adopt the virial temperature for the smallest
minihalos that can cool via ${\rm H_2}$ and form stars to be $T_{\rm
min}=400$K (Haiman et al. 1996; Tegmark et al. 2001; Machacek et
al. 2001 -- note that the choice in HH03 was lower, $T_{\rm min}=100$
K).

Gas infall is expected to be suppressed for sufficiently small halos
in actively ionized and photo--heated patches of the IGM.  We define a
temperature $T_{\rm uv}$ and only permit halos with virial
temperatures below this value to form ionizing sources in the neutral
fraction $1-F_{\rm HII}(z)$ of the volume of the IGM. The value of
$T_{\rm uv}$ is uncertain -- at low redshifts it is commonly taken as
$\approx 2.5 \times 10^5$ K, corresponding to a velocity dispersion
$\sigma = 50 {\rm~km~s^{-1}}$ (e.g., Thoul \& Weinberg 1996).
However, Dijkstra et al. (2004) showed that this suppression is less
efficient at high redshift, and based on their result for halos
collapsing at redshifts $6\lsim z\lsim 10$, we adopt the fiducial
value of $T_{\rm uv}=9\times 10^4$K (or $\sigma \sim 30 {\rm
km~s^{-1}}$).

In order to match inferences from the $z\lsim 6.5$ quasar spectra
(Becker et al. 2001; Fan et al. 2006), we require that percolation
takes place at redshift $z = 6.5$.  At this redshift, ionization is
dominated by larger halos, and so this requirement effectively fixes
the value of $\epsilon_{\rm large}=120$.\footnote{ HH03 found a value
that was 50\% smaller -- we here find that the efficiency increases by
a factor of 3 because of the lower small scale power implied by the
3--yr {\it WMAP} data (lower $\sigma_8$ and $n_s$), but was decreased
by a factor of 2 due to our lower value of $T_{\rm uv}$.}  This value
of the overall efficiency is reasonable: for a normal stellar
population with a Salpeter mass function, $N_\gamma=4000$, and we may
break down the efficiency into $f_*=0.15$ and $f_{\rm esc}=0.2$.  The
star--formation rate at $z\sim 6$ predicted with these assumptions is
consistent with the global value inferred from the Hubble Ultra Deep
Field (see Mesinger et al. 2005 for a comparison).

\begin{figure}[t]
\plotone{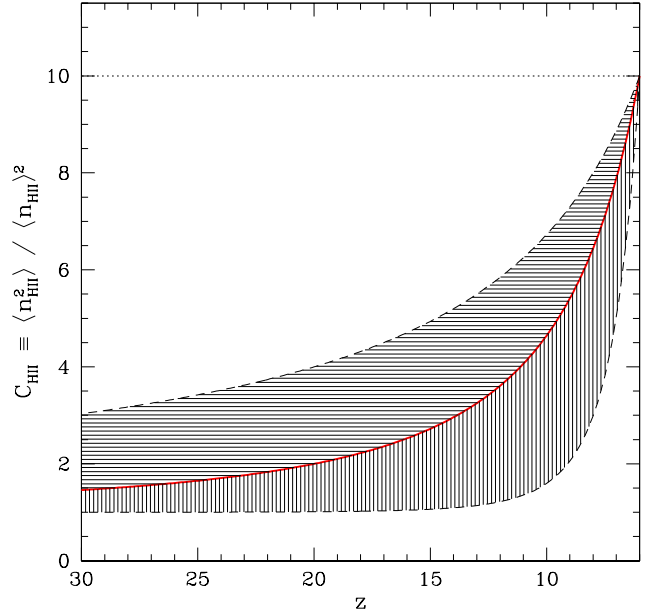}
\caption{The assumed evolution of the clumping factor of ionized gas
with redshift.  The thick solid (red) curve is adopted as our fiducial
case, and has a decrease by a factor of $\sim 4$ between $z=6$ and
$z=15$.  The shaded range indicates models in which this decrease
varies from a factor of $2-10$; the limiting case of a constant
clumping factor of $C_{\rm HII}=10$ is shown as the dotted line.}
\label{fig:clumping} 
\end{figure}

Finally, the evolution of the ionized fraction depends on the clumping
factor of the ionized gas, $C_{\rm HII}\equiv \langle n_{\rm
HII}^2\rangle / \langle n_{\rm HII}\rangle^2$.  The value of the
clumping factor at $z\sim 6$ does not impact our results because it is
absorbed into $\epsilon_{\rm large}$, which, in turn, is set by the
requirement that the ionization percolates at $z=6.5$.  This
normalization requires $\epsilon_{\rm large}\approx 120\times C_{\rm
HII}(z=6)/10$. However, we find that the {\it evolution} of $C_{\rm
HII}$ with redshift has a significant effect on our results.  We do
not attempt to model this evolution ab--initio; instead we adopt a
relation that is a conservative compromise between three different
ab--initio approaches, and we vary our parameters to reflect the range
of predictions of the three models (see discussion in
\S~\ref{sec:fiducial} below).  The relation we adopt is a simple
power-law,
\beq
C_{\rm HII}(z)-1 = 9\left(\frac{1+z}{7}\right)^{-\beta},
\label{eq:clumping}
\eeq
which has a fixed $C_{\rm HII}(z=6)=10$, and declines towards high
redshift at a rate given by the slope $\beta$.  What matters most for
our results below is the decrease between $z=6$ (the redshift which
sets $\epsilon_{\rm large}$) and $15\lsim z\lsim 20$ (the redshift
which determines the contribution to $\tau$ from the minihalos).  In
our fiducial model, $\beta=2$, corresponding to the ratio $C_{\rm
HII}(z=6)/C_{\rm HII}(z=15)=3.7$. This choice, together with the range
$1\leq \beta\leq 6$ (corresponding to a range of ratios $2< C_{\rm
HII}(z=6)/C_{\rm HII}(z=15)< 10$), is shown in
Figure~\ref{fig:clumping}.  For reference, the dotted line shows the
limiting case of a constant clumping factor ($\beta=0$).

In summary, our models have only two free parameters: $\epsilon_{\rm
mini}$, and $\beta$.  In addition, we will consider uncertainties in
$T_{\rm min}$, the minimum mass for efficient ${\rm H_2}$ cooling, and
$T_{\rm uv}$, the maximum virial temperature of halos that can collect
photo--heated gas.  We next discuss our choices for the fiducial
values of these parameters.

\begin{figure}[t]
\plotone{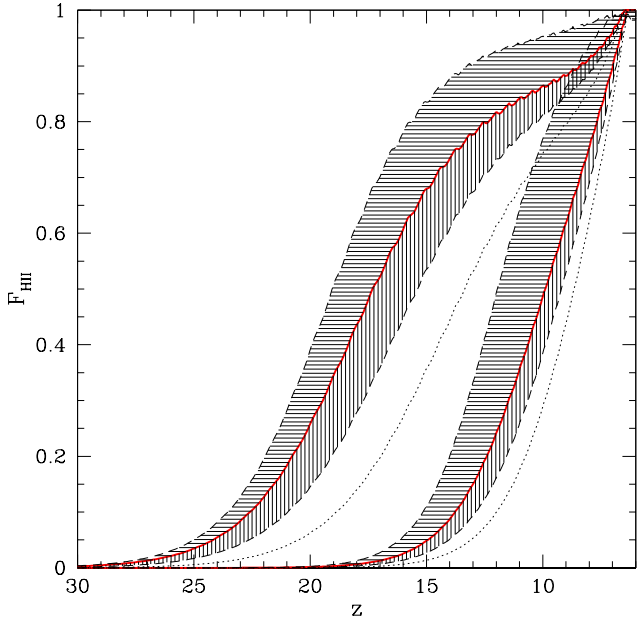}
\caption{The evolution of the ionized fraction of hydrogen.  The thick
solid (red) curves, the shaded ranges, and the limiting cases shown by
the dotted curves, correspond to the different logarithmic slopes for
the evolution of the clumping factor with redshift $C_{\rm HII}(z)$
shown in Figure~\ref{fig:clumping}.  The lower set of four curves
assume star--formation in minihalos (with $400{\rm K}<T_{\rm
vir}<10^4{\rm K}$) has been completely suppressed ($\epsilon_{\rm
mini}=0$), and the upper set of curves assumes minihalos inject
ionizing photons into the IGM with the fiducial efficiency
($\epsilon_{\rm mini}=200$).}
\label{fig:xe} 
\end{figure}

\section{Fiducial Parameters}
\label{sec:fiducial}

What is the expected value of $\epsilon_{\rm mini}$?  Abel, Bryan, \&
Norman (2002), and Bromm, Coppi \& Larson (2002) suggest that
minihalos form metal--free stars from only a small fraction
($f_*\approx 0.0025$) of the available gas. On the other hand, the
stars are, on average, $\approx 20$ times more efficient ionizing
photon producers (per unit stellar mass) because they are massive
($M\gsim 100~{\rm M_\odot}$) and have harder spectra (Tumlinson \&
Shull 2000; Bromm, Kudritzki \& Loeb 2001; Schaerer 2002). As a
result, $N_\gamma\approx 80000$; and all of their ionizing radiation
escapes into the IGM, at least from the smaller minihalos that
contribute most of the high--redshift tail for reionization ($f_{\rm
esc}=1$; Whalen, Abel \& Norman 2004). This results in the overall
fiducial efficiency of $\epsilon_{\rm mini}=\epsilon_{\rm fid}=200$.
We note that this is quite close to the value $\epsilon_{\rm
large}=120$ that we inferred for the large halos by requiring
percolation to occur at $z\approx 6.5$.

Several authors have attempted to compute the clumping factor $C_{\rm
HII}(z)$ from first principles.  In the simulations of Gnedin \&
Ostriker (1997), it increases dramatically (by a factor of $\sim 50$)
between $z=6$ and $z=15$, but since they do not resolve small-scale
fluctuations early on, this ratio may be an overestimate. Iliev,
Scannapieco \& Shapiro (2005) have a different approach of separately
estimating the contribution from the low--density IGM from a
simulation, and adding to this semi-analytic estimates of the clumping
due to minihalos (Haiman, Abel \& Madau 2001).  They find an evolution
that is less steep, with $C_{\rm HII}(z=10)/C_{\rm HII}(z=15)\approx
2$, but they do not cover the interesting $z<10$ range, where we
normalize our models, and where $C_{\rm HII}$ likely increases
rapidly.  Finally, Miralda-Escud\'e, Haehnelt \& Rees (2000) offer a
different prescription: they assume that the IGM is ionized
``outside--in'', with all the gas below a critical overdensity
$\Delta_{\rm crit}$ ionized at a given redshift (and neutral above
this threshold), to compute $C_{\rm HII}$ directly. This prescription
predicts relatively slow evolution, with $\Delta_{\rm
crit}\approx10-20$, until the late stages of reionization (as in
Wyithe \& Loeb 2003 and Wyithe \& Cen 2006). However, towards the end
of the reionization epoch, $\Delta_{\rm crit}$ increases rapidly. This
is an important distinction for our present purposes, because the
ionizing emissivity in our models is normalized at the end--stages of
reionization.  Miralda-Escud\'e, Haehnelt \& Rees (2000) show that by
the time the mean free path of ionizing photons grows to the Hubble
length (which appears to be the case at $z\sim 6$; Lidz et al. 2005),
$\Delta_{\rm crit}\gsim 500$, which will correspond to a steep rise in
the clumping factor.

\begin{figure}[t]
\plotone{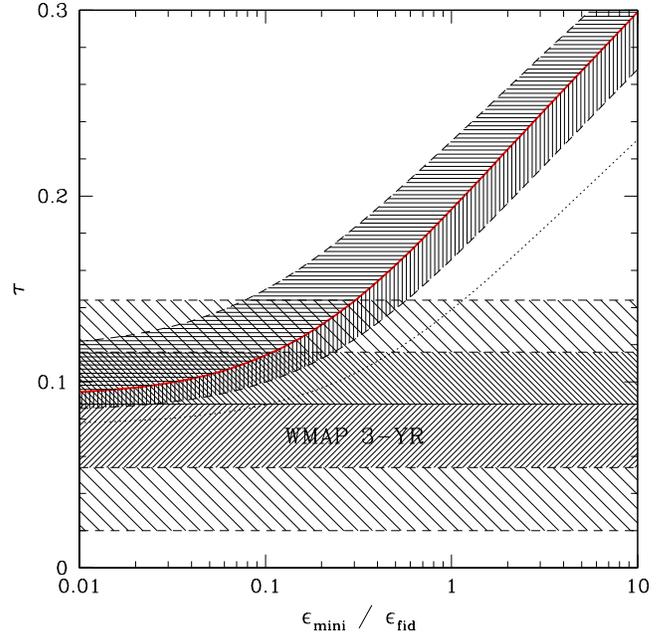}
\caption{The optical depth to electron scattering $\tau$, as a
function of the efficiency of minihalos to inject ionizing photons
into the IGM.  The efficiency is shown in units of the fiducial value
$\epsilon_{\rm fid}=200$ (see text for discussion).  The thick solid
(red) curve, the shaded range, and the limiting case shown by the
dotted curve, follow the notation for models with different clumping
factor evolutions shown in Figures~\ref{fig:clumping} and
\ref{fig:xe}.  The horizontal lanes show the range of $\tau$ allowed
by the {\it WMAP} three--year data (at 68\% and 95\% CL).  }
\label{fig:taus} 
\end{figure}

\section{Results and Discussion}
\label{sec:results}

The evolution of the ionized fraction of hydrogen in several models is
shown in Figure~\ref{fig:xe}.  The thick solid (red) curves, the
shaded ranges, and the limiting cases shown by the dotted curves,
correspond to the different clumping factor evolutions $C_{\rm
HII}(z)$ shown in Figure~\ref{fig:clumping} with the same notation.
The lower set of curves assumes star--formation in minihalos has been
completely suppressed ($\epsilon_{\rm mini}=0$), and the upper set of
curves assumes minihalos inject ionizing photons into the IGM with the
fiducial efficiency ($\epsilon_{\rm mini}=200$).  All curves are
normalized to produce full reionization at $z\sim 6.5$. Earlier
reionization is consistent with SDSS quasar spectra; this would yield
a larger $\tau$, and would strengthen our conclusions.

The models shown in Figure~\ref{fig:xe} that include minihalos have
optical depths of $\tau=0.14, 0.17, 0.19$, and $0.23$, and are all
ruled out by the three--year {\it WMAP} data at more than 95\% CL.
The models that exclude minihalos have $\tau=$ 0.077, 0.084, 0.091,
and 0.12, and all except the case with the steepest $C_{\rm HII}(z)$
evolution are within $1\sigma$ of the {\it WMAP} value.

In Figure~\ref{fig:taus}, we show $\tau$ as a function of the
efficiency $\epsilon_{\rm mini}$.  The figure follows the notation of
Figures~\ref{fig:clumping} and \ref{fig:xe} to describe the four
different cases for the evolution of the clumping factor.  The main
conclusion that can be drawn from this figure is that the photon
production efficiency in minihalos was smaller than the expected
fiducial value by a factor of $\gsim 2-10$.  This also means that
minihalos contribute to reionization with a {\it lower} efficiency
than the larger halos.

\begin{figure}[t]
\plotone{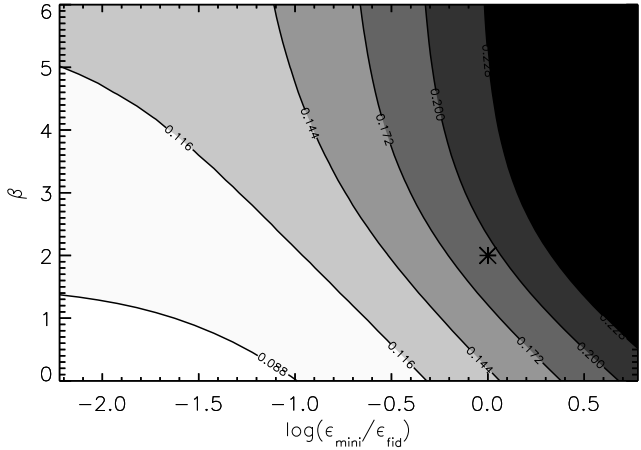}
\caption{The predicted optical depth as a function of the ionizing
efficiency of the minihalos ($\epsilon_{\rm mini}/\epsilon_{\rm fid}$)
and the logarithmic slope of the decline of the clumping factor with
redshift ($\beta\equiv d\log(C_{\rm HII}-1)/d\log(1+z)$). The lowest
contour (white region) corresponds to optical depths within one sigma
of the 3--yr {\it WMAP} result, and each contour is a one-sigma
increase in the optical depth.  The fiducial model is marked with an
asterisk and deviates from the {\it WMAP} value by nearly $4\sigma$.}
\label{fig:contour}
\end{figure}

More generally, Figure~\ref{fig:contour} shows the predicted optical
depth as a function of both $\epsilon_{\rm mini}$ and $\beta$. The
lowest contour (white region) corresponds to optical depths within
1$\sigma$ of the 3--yr {\it WMAP} value, and each successive contour
shows a 1$\sigma$ increase.  The figure shows that our fiducial model
deviates from the {\it WMAP} value by nearly $4\sigma$; consistency at
the $1-2\sigma$ level requires either reducing $\epsilon_{\rm mini}$
by a factor of 2--10, or adopting an essentially non--evolving
clumping factor ($\beta\approx 0$) and a smaller reduction in
$\epsilon_{\rm mini}$.

In order to examine the robustness of our conclusions to uncertainties
in the model, we have repeated our calculations for different values
of $T_{\rm min}$, the virial temperature of the smallest minihalo, and
$T_{\rm uv}$, the critical virial temperature for suppression of gas
infall by photoionization heating.  We find that variations in the
range $160~{\rm K}\leq T_{\rm min} \leq 10^3~{\rm K}$ and $10^4~{\rm
K}\leq T_{\rm uv} \leq 2\times 10^5~{\rm K}$ change the required
reduction factor of $\epsilon_{\rm mini}$ by less than 30\%.

We emphasize that our predicted range of $\tau=0.08-0.11$, for models
that exclude minihalos, is consistent with numerous previous
semi--analytic studies (as well as simulations that do not resolve
minihalos, e.g., Ciardi et al. 2003), in the wake of the 1--yr
announcement of the {\it WMAP} measurements. In another recent study,
Wyithe \& Cen (2006) examined populations of metal--free stars in
$T_{\rm vir}\gsim 2\times10^4$K halos, with an efficiency similar to
normal stars, and found similar $\tau$ values.  It is also worth
noting that the modified Press--Schechter form for the DM halo mass
function that we used (Jenkins et al. 2001) has been directly probed
and confirmed in the relevant mass and redshift range by recent
three--dimensional simulations (Yoshida et al. 2003; Jang-Condell \&
Hernquist 2001; Springel et al. 2005).  However, the minihalos arise
from primordial perturbations on very small ($k\sim 10-100$) scales,
and predicting their abundance does involve an extrapolation of the
measured power spectrum by $\sim 2-3$ orders of magnitude in $k$
(Barkana et al. 2001; Mesinger et al. 2005).

\newpage
\section{Conclusions}
\label{sec:conclusions}

The {\it WMAP} experiment has opened a new window into studies of the
first structures at the end of the cosmological dark ages.  The
first--year data release has suggested the large optical depth
$\tau\sim 0.17$ (Spergel et al. 2003), which appeared in conflict with
simple predictions, and resulted in different suggestions to account
for this large $\tau$.  We argue here that the significantly lower
$\tau\approx 0.09\pm 0.03$ allowed by the three--year data is once
again interesting: if minihalos were forming stars efficiently, the
expected value would be higher, $\tau\gsim 0.17$.  To reproduce the
new value for $\tau$ requires an order of magnitude reduction in the
fiducial efficiency of ionizing photon production in minihalos.

Star--formation in minihalos has been predicted theoretically to be
significantly suppressed, and the new {\it WMAP} data has provided the
first interesting empirical evidence for this suppression.  In
addition to the various physical feedback effects mentioned in the
Introduction, the clustering of the early minihalos will, in general,
help suppressing more halos at high redshift, since clustering will
place more minihalos in the destructive sphere of influence of earlier
collapsed objects. In the presence of clustering, photoionization
heating alone may account for the low $\tau$ (Kramer et al. 2006).

If massive Pop III stars typically leave behind intermediate--mass
black holes, these may accrete efficiently and produce more
significant ionization than their progenitor stars (Madau et al. 2004;
Ricotti \& Ostriker 2004).  Our results, by inference, provides
similarly interesting constraints on this scenario.

Further studies from CMB anisotropies, spectroscopic observations of
quasars and Ly$\alpha$ emitting galaxies at $z>6$, radio probes of the
redshifted 21cm line of neutral hydrogen, as well as direct detections
of the sources of reionization and their end--products as supernovae
and gamma ray bursts, will clarify the star--formation efficiency in
minihalos and their contribution to reionization.

\acknowledgements{ZH acknowledges partial support by NASA through
grants NNG04GI88G and NNG05GF14G, by the NSF through grants
AST-0307291 and AST-0307200, and by the Hungarian Ministry of
Education through a Gy\"orgy B\'ek\'esy Fellowship.  GB acknowledges
partial support by NSF grants AST-0507161 and AST-0547823.}

\end{document}